\documentclass[aps,reprint]{revtex4-2} 
\usepackage{graphicx,graphics} 
\usepackage{epstopdf} 
\usepackage{dcolumn}
\usepackage{amsmath,amssymb,amsfonts} 
\usepackage{latexsym,verbatim} 
\usepackage{bm} 
\usepackage{bbold} 
\usepackage{xcolor} 
\usepackage{ulem} 
\usepackage[breaklinks=false,colorlinks,citecolor=blue,linkcolor=blue,urlcolor=blue]{hyperref} 
\usepackage[abs]{overpic} 
\usepackage{textcomp} 
\usepackage{mathtools} 
\usepackage{braket} 
\usepackage{soul} 
\usepackage{lipsum}
\usepackage[export]{adjustbox}
\usepackage{gensymb}
\usepackage[T1]{fontenc}
\usepackage{multirow}
\usepackage{tikz}

\begin{document}

\title{High-fidelity controlled-phase gates for distinguishable quantum walkers via extended interactions}

\author{Gaia Forghieri$^{1}$}
\email{gaia.forghieri@unimi.it}
\author{Matteo G. A. Paris$^{1}$}
\affiliation{$^{1}$Dipartimento di Fisica, Università di Milano, I-20133 Milan, Italy}

\date{\today}

\begin{abstract}
We investigate the implementation of controlled-phase (CP) gates with quantum walks in a dual-rail encoding through the use of interacting particles. While previous proposals have focused on indistinguishable particles (bosons or fermions) to achieve unitary fidelity for plane-wave scattering, practical implementations require finite-size wavepackets and routing through single-particle gates, both factors that introduce unavoidable fidelity losses. We show that extending the interaction range beyond on-site or first-neighbor terms provides sufficient control over the scattering potential to engineer CP gates with distinguishable particles that match the ideal bosonic/fermionic performance. For finite Gaussian wavepackets, we derive analytical approximations for the gate fidelity in terms of the transmission coefficient's magnitude and phase derivatives. We find that while distinguishable-particle scattering alone exhibits slightly lower fidelity than the indistinguishable case, the overall architecture that we propose avoids the additional single-particle gates required for routing indistinguishable particles. The roundabout gates needed for the latter indeed introduce fidelity losses approximately one order of magnitude larger than the interaction-induced losses, making the distinguishable-particle approach competitive for practical implementations. Our results establish multi-neighbor interactions as a tool for quantum information processing with continuous-time quantum walks and provide quantitative guidelines for optimizing gate fidelities in finite-size systems.
\end{abstract}

\maketitle

\section{Introduction}

{Continuous-time quantum walks have emerged as a powerful and versatile framework for quantum information processing, providing a natural setting in which transport \cite{mulken2011,annoni2024,razzoli2021,finocchiaro2025}, entanglement generation \cite{forghieri2026,wang2014,preiss2015,tziperman2026,benedetti2012}, and quantum logic operations \cite{lahini2018,chapman2024,hassani25,herrman2019} can be described within a unified dynamical picture. Over the last two decades, quantum walks have been widely investigated both from the algorithmic perspective and as physical platforms for quantum technologies \cite{ambainis2003,yan2019,qiang2021,venegas-andraca2012,grafe2016,kadian2021}. This effort has led to applications ranging from quantum search optimization \cite{candeloro2023,apers2022,lugao2024} and perfect routing protocols \cite{bottarelli2023,cavazzoni2022,ragazzi2025} to the analysis of complex networks \cite{magano2023,forghieri2026-linkprediction,faccin2014}, to universal models for quantum computation \cite{underwood2012, qiang2024,gonzales2025}. Specifically, the combination of a dual-rail encoding with interacting quantum walkers offers a promising route to implementing quantum circuits in which the state of a qubit is represented by the path of a quantum walker \cite{childs2009, childs2013, asaka2023, guan2026}. In this architecture, single-qubit gates are realized through single-particle scattering processes, and entangling gates arise from controlled interactions between multiple walkers. Within this framework, controlled-phase (CP) gates play a central role, as they provide the nontrivial two-qubit operation required for universal quantum computation \cite{nielsen2010,bremner2002}. Several proposals for such CP gates have primarily focused on indistinguishable particles (bosons or fermions), exploiting their exchange statistics to achieve perfect, unitary fidelity in the ideal limit of plane-wave scattering \cite{childs2013,asaka2023}. However, a significant challenge arises when moving from these idealized descriptions to practical implementations \cite{wang2013physical,qiang2024}. In this context indeed, one needs to account for wavepacket dispersion, momentum-dependent scattering phases, and the requirement for additional single-particle routing operations. All of these factors inevitably introduce fidelity losses that may compromise the performance of the circuit. Understanding how these effects can be mitigated while preserving the advantages of quantum-walk-based computation is therefore a key challenge for the development of scalable architectures.}


In this work, we address the above-mentioned limitations by proposing and analyzing an alternative architecture for CP gate implementation that utilizes distinguishable particles. Our approach departs from previous schemes by eliminating the need for complex routing gates that redirect  particles onto a common path to make them indistinguishable \cite{childs2013,asaka2023}. Instead, we employ distinguishable walkers that propagate on separate, parallel chains and interact through a distance-dependent potential. {In platforms such as coupled waveguides \cite{rai2008,zhou2024,gao2024,raymond25}, trapped-ion arrays \cite{zahringer2010,tamura2020,huerta-alderete2020}, or Rydberg lattices \cite{cote2006,khazali2022,chen2024,palaiodimopoulos2024}, this effective interaction can be engineered via inter-lattice distances and potential shapes.} Consequently, we demonstrate that by extending the interaction range beyond the typical on-site or nearest-neighbor terms, we gain sufficient control over the scattering potential to engineer CP gates with distinguishable particles that match the ideal performance of their indistinguishable counterparts for plane waves. Crucially, by avoiding the single-particle routing gates, our proposed architecture significantly reduces the overall fidelity loss in the presence of finite-sized wavepackets. Through analytical approximations and numerical simulations, we show that the fidelity loss introduced by the roundabout gates in the indistinguishable-particle architecture is approximately one order of magnitude larger than the interaction-induced losses for the distinguishable case. This makes our distinguishable-particle approach highly competitive and more practical for experimental realizations, establishing multi-neighbor interactions as a key tool for robust quantum information processing with continuous-time quantum walks.

\begin{figure*}[t]
    \centering
  \begin{minipage}{0.495\textwidth}
    \centering
    \resizebox{\textwidth}{!}{\usetikzlibrary{arrows.meta, positioning}

\begin{tikzpicture}[
    font=\large,
    scale=1,
    line width=0.8pt,
    walker/.style={circle, draw, fill=black, inner sep=1.5pt},
    emptydot/.style={circle, draw, inner sep=1pt,fill=black},
    bigwalker1/.style={circle, draw, minimum size=4mm,fill=red,opacity=0.3},
    bigwalker2/.style={circle, draw, minimum size=4mm,fill=red,opacity=1},
    phasebox/.style={rectangle, draw, minimum width=1.3cm, minimum height=2.6cm, fill=blue,opacity=0.4},
    roundabout/.style={circle, draw, minimum width=1.25cm}
]



\def\yA{5}   
\def\yB{4}   
\def\yC{1}   
\def\yD{0}   
\def\dy{0.2}


\node[left] at (-0.5,5.2) {\Large (a)};

\node[left] at (0.3,\yA) {$|0\rangle$};
\node[left] at (0.3,\yB) {$|1\rangle$};
\node[left] at (0.3,\yC) {$|0\rangle$};
\node[left] at (0.3,\yD) {$|1\rangle$};

\node[left] at (-0.3,0.5*\yA+0.5*\yB) {$Q_1$};
\node[left] at (-0.3,0.5*\yC+0.5*\yD) {$Q_2$};

\node[right] at (1.3,4.5) {in};
\node[right] at (1.3,0.5) {in};

\node[right] at (10.3,4.5) {out};
\node[right] at (10.3,0.5) {out};


\node[bigwalker1] at (2,\yA) {};
\node[bigwalker1] at (2,\yB) {};

\node[bigwalker2] at (2,\yD) {};

\draw[->] (2.2,\yA+\dy) -- +(0.4,0);
\draw[->] (2.2,\yB+\dy) -- +(0.4,0);
\draw[->] (2.2,\yD+\dy) -- +(0.4,0);


\node[bigwalker1] at (4.4,2) {};
\node[bigwalker2] at (4.6,2) {};

\draw[->] (4.2,1.8) -- +(0,-0.4);
\draw[->] (4.8,2.2) -- +(0,0.4);

\node[bigwalker1,fill=blue] at (7,0.18) {};
\node[bigwalker2,fill=blue] at (7,3.82) {};

\draw[->] (7.,0.5) -- +(0.1,0.4);
\draw[->] (7.,3.5) -- +(0.1,-0.4);


\node[bigwalker1] at (10,\yA) {};
\node[bigwalker1, fill=blue] at (10,\yB) {};
\node[bigwalker2, fill=blue] at (10,\yD) {};

\draw[->] (10.2,\yA+\dy) -- +(0.4,0);
\draw[->] (10.2,\yB+\dy) -- +(0.4,0);
\draw[->] (10.2,\yD+\dy) -- +(0.4,0);


\draw (0.8,\yA) -- (5.2,\yA);
\draw (6.8,\yA) -- (11.2,\yA);

\draw (0.8,\yB) -- (4,\yB);
\draw (5,\yB) -- (6.5,\yB);
\draw (8.5,\yB) -- (11.2,\yB);

\draw (0.8,\yC) -- (3.2,\yC);
\draw (8.8,\yC) -- (11.2,\yC);

\draw (0.8,\yD) -- (4,\yD);
\draw (5,\yD) -- (6.5,\yD);
\draw (8.5,\yD) -- (11.2,\yD);

\node[right] at (11.1,\yA) {$\cdots$};
\node[right] at (11.1,\yB) {$\cdots$};
\node[right] at (11.1,\yC) {$\cdots$};
\node[right] at (11.1,\yD) {$\cdots$};

\node[left] at (0.9,\yA) {$\cdots$};
\node[left] at (0.9,\yB) {$\cdots$};
\node[left] at (0.9,\yC) {$\cdots$};
\node[left] at (0.9,\yD) {$\cdots$};

\node[right] at (3.1,\yC) {$\cdots$};
\node[left]  at (8.9,\yC) {$\cdots$};
\node[right] at (5.1,\yA) {$\cdots$};
\node[left]  at (6.9,\yA) {$\cdots$};

\foreach \x in {1,2,3,4,5,7,8,9,10,11} {
    \node[emptydot] at (\x,\yA) {};
}
\foreach \x in {1,2,3,9,10,11} {
    \node[emptydot] at (\x,\yC) {};
}
\foreach \x in {1,2,3,4,5,6,9,10,11} {
    \node[emptydot] at (\x,\yB) {};
}
\foreach \x in {1,2,3,4,5,6,9,10,11} {
    \node[emptydot] at (\x,\yD) {};
}

\node[emptydot] at (7,0.18) {};
\node[emptydot] at (7.42,1.1) {};
\node[emptydot] at (7.58,2.9) {};
\node[emptydot] at (8,3.82) {};

\node[emptydot] at (7,3.82) {};
\node[emptydot] at (7.42,2.9) {};
\node[emptydot] at (7.58,1.1) {};
\node[emptydot] at (8,0.18) {};




\node[phasebox] at (4.5,0.5*\yB+0.5*\yD) {};
\node[font=\Large] at (3.5,2.8) {$\theta$};

\node[roundabout] at (4.5,3.625) {};
\node[roundabout] at (4.5,0.375) {};
\draw (4.5,1) -- (4.5,3);

\node[emptydot] at (4.5,1) {};
\node[emptydot] at (4.5,2) {};
\node[emptydot] at (4.5,3) {};

\draw [->] (4.8,0.05) to [out=210,in=-30] (4.2,0.05);
\draw [->] (4.15,0.15) to [out=120,in=180] (4.45,0.75);
\draw [->] (4.55,0.75) to [out=0,in=60] (4.85,0.15);

\draw [->] (4.8,3.95) to [out=150,in=30] (4.2,3.95);
\draw [->] (4.15,3.85) to [out=240,in=180] (4.45,3.25);
\draw [->] (4.55,3.25) to [out=0,in=-60] (4.85,3.85);

\draw (6.5,0) to [out=0,in=180] (8.5,4);
\draw (6.5,4) to [out=0,in=180] (8.5,0);

\end{tikzpicture}}
  \end{minipage}
  \hfill
  \begin{minipage}{0.495\textwidth}
    \centering
    \resizebox{\textwidth}{!}{\usetikzlibrary{arrows.meta, positioning}

\begin{tikzpicture}[
    font=\large,
    scale=1,
    line width=0.8pt,
    walker/.style={circle, draw, fill=black, inner sep=1.5pt},
    emptydot/.style={circle, draw, inner sep=1pt,fill=black},
    bigwalker1/.style={circle, draw, minimum size=4mm,fill=red,opacity=0.3},
    bigwalker2/.style={circle, draw, minimum size=4mm,fill=red,opacity=1},
    phasebox/.style={rectangle, draw, minimum width=1.6cm, minimum height=2.6cm, fill=blue,opacity=0.4}
]



\def\yA{5}   
\def\yB{4}   
\def\yC{1}   
\def\yD{0}   
\def\dy{0.2}


\node[left] at (-0.5,5.2) {\Large (b)};

\node[left] at (0.3,\yA) {$|0\rangle$};
\node[left] at (0.3,\yB) {$|1\rangle$};
\node[left] at (0.3,\yC) {$|0\rangle$};
\node[left] at (0.3,\yD) {$|1\rangle$};

\node[left] at (-0.3,0.5*\yA+0.5*\yB) {$Q_1$};
\node[left] at (-0.3,0.5*\yC+0.5*\yD) {$Q_2$};

\node[right] at (1.3,4.5) {in};
\node[right] at (1.3,0.5) {in};

\node[right] at (10.3,4.5) {out};
\node[right] at (10.3,0.5) {out};


\node[bigwalker1] at (2,\yA) {};
\node[bigwalker1] at (2,\yB) {};

\node[bigwalker2] at (2,\yD) {};

\draw[->] (2.2,\yA+\dy) -- +(0.4,0);
\draw[->] (2.2,\yB+\dy) -- +(0.4,0);
\draw[->] (2.2,\yD+\dy) -- +(0.4,0);


\node[bigwalker1] at (4,2) {};
\node[bigwalker2] at (5,2) {};

\draw[->] (4.2,1.8) -- +(0,-0.4);
\draw[->] (4.8,2.2) -- +(0,0.4);

\node[bigwalker1,fill=blue] at (7,0.38) {};
\node[bigwalker2,fill=blue] at (7,3.82) {};

\draw[->] (7.,0.7) -- +(0.1,0.6);
\draw[->] (7.,3.5) -- +(0.1,-0.4);




\node[bigwalker1] at (10,\yA) {};
\node[bigwalker1, fill=blue] at (10,\yB) {};
\node[bigwalker2, fill=blue] at (10,\yD) {};

\draw[->] (10.2,\yA+\dy) -- +(0.4,0);
\draw[->] (10.2,\yB+\dy) -- +(0.4,0);
\draw[->] (10.2,\yD+\dy) -- +(0.4,0);


\draw (0.8,\yA) -- (5.2,\yA);
\draw (6.8,\yA) -- (11.2,\yA);

\draw (0.8,\yB) -- (4,\yB);
\draw (4,\yB) -- (4,\yD+\dy);
\draw (4,\yD+\dy) -- (5-\dy,\yD+\dy);
\draw (5+\dy,\yD+\dy) -- (6.5,\yD+\dy);
\draw (8.5,\yB) -- (11.2,\yB);

\draw (0.8,\yC) -- (3.2,\yC);
\draw (8.8,\yC) -- (11.2,\yC);

\draw (0.8,\yD) -- (5,\yD);
\draw (5,\yD) -- (5,\yB);
\draw (5,\yB) -- (6.5,\yB);
\draw (8.5,\yD) -- (11.2,\yD);

\node[right] at (11.1,\yA) {$\cdots$};
\node[right] at (11.1,\yB) {$\cdots$};
\node[right] at (11.1,\yC) {$\cdots$};
\node[right] at (11.1,\yD) {$\cdots$};

\node[left] at (0.9,\yA) {$\cdots$};
\node[left] at (0.9,\yB) {$\cdots$};
\node[left] at (0.9,\yC) {$\cdots$};
\node[left] at (0.9,\yD) {$\cdots$};

\node[right] at (3.1,\yC) {$\cdots$};
\node[left]  at (8.9,\yC) {$\cdots$};
\node[right] at (5.1,\yA) {$\cdots$};
\node[left]  at (6.9,\yA) {$\cdots$};

\foreach \x in {1,2,3,4,5,7,8,9,10,11} {
    \node[emptydot] at (\x,\yA) {};
}
\foreach \x in {1,2,3,4,5,9,10,11} {
    \node[emptydot] at (\x,\yA) {};
    \node[emptydot] at (\x,\yC) {};
}
\foreach \x in {1,2,3,4,5,6,9,10,11} {
    \node[emptydot] at (\x,\yB) {};
}
\foreach \x in {1,2,3,4,5,9,10,11} {
    \node[emptydot] at (\x,\yD) {};
}
\foreach \x in {4,5+\dy,6+0.5*\dy} {
    \node[emptydot] at (\x,\yD+\dy) {};
}

\node[emptydot] at (7,0.38) {};
\node[emptydot] at (7.35,1.1) {};
\node[emptydot] at (7.58,2.9) {};
\node[emptydot] at (8,3.82) {};

\node[emptydot] at (7,3.82) {};
\node[emptydot] at (7.42,2.9) {};
\node[emptydot] at (7.58,1.1) {};
\node[emptydot] at (8,0.18) {};

\begin{scope}
    \clip (4.5,0.185) rectangle (5.5,0.5);
    \draw (5,0.2) circle(0.19);
\end{scope}



\node[phasebox] at (4.5,0.5*\yB+0.5*\yD) {};
\node[font=\Large] at (3.35,2.8) {$\theta$};


\node[emptydot] at (4,2) {};
\node[emptydot] at (4,3) {};
\node[emptydot] at (5,2) {};
\node[emptydot] at (5,3) {};



\draw (6.5,\dy) to [out=0,in=180] (8.5,4);
\draw (6.5,4) to [out=0,in=180] (8.5,0);

\end{tikzpicture}}
  \end{minipage}
  \caption{Dual-rail implementations of a CP gate based on interacting quantum walkers, as described in the main text. In both panels, the logical state $|1\rangle$ of each qubit is routed through an interaction region (blue box), where the walkers acquire a conditional phase $\theta$. (a) Implementation based on indistinguishable particles, following Ref. \cite{asaka2023}, where roundabout gates route the walkers onto a common path where the walkers can interact, and restore the original encoding afterwards. (b) Alternative implementation based on distinguishable particles propagating on parallel chains and interacting through a distance-dependent potential, eliminating the need for additional routing gates.}
  \label{fig:fig1}
\end{figure*}
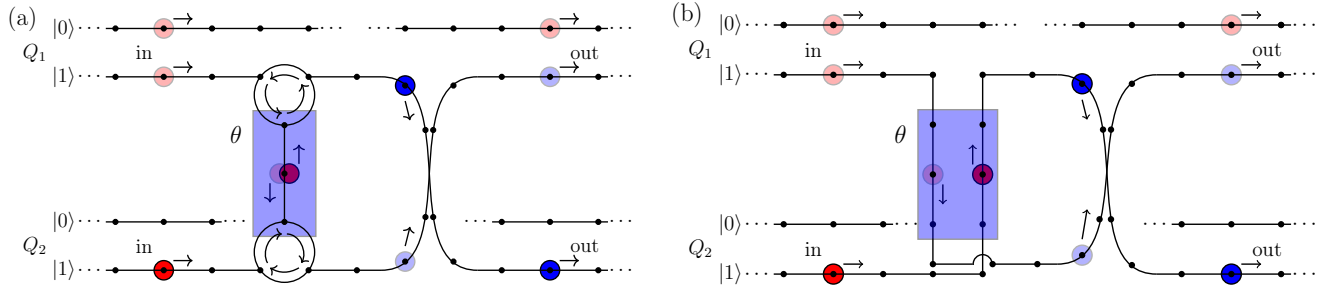

This paper is structured as follows. In Sec. \ref{sec:model} we describe the studied system, starting from a comparison between the indistinguishable- vs. distinguishable-particle architectures for CP gates in Sec. \ref{sec:model_circuit}. We then introduce the mathematical description of the scattering theory for interacting quantum walks in Sec. \ref{sec:walker_theory}, and define Gaussian wavepackets in Sec. \ref{sec:wavepacket_theory}. We report the results on gate fidelities in Sec. \ref{sec:fidelity_results}. We first report in Sec. \ref{sec:ideal_fidelity} an analysis of the impact of multi-neighbor interactions on the distinguishable-particle fidelity of the ideal, plane-wave implementation, and compare the performance with the indistinguishable case. Then, we introduce finite-size effects in Sec. \ref{sec:finite_fidelity}, and compare the fidelity loss between the indistinguishable- and distinguishable- particle implementations. We provide both numerical calculations of the wavepacket fidelity through the time evolution of the interacting particles, and analytical approximations in the limit of narrow momentum distributions. Sec. \ref{sec:conclusions} summarizes our conclusions. Lastly, we explicitely report the solutions for the plane-wave scattering coefficients in Appendix \ref{app:scattering}, and show the derivation for the analytical approximations of the fidelity with Gaussian wavepackets in Appendix \ref{app:finite_fidelity}.

\section{Model system}\label{sec:model}

\subsection{Description of the circuit}
\label{sec:model_circuit}

In a dual-rail architecture, the state of a single qubit is encoded by the presence of a walker in one of two parallel paths, so that an $n$-qubit state is encoded by the positions of $n$ walkers on $2n$ paths, as schematically illustrated in Fig. \ref{fig:fig1}. Within this framework, each walker is initialized on the left, as a superposition of states in the upper and lower rails representing kets $|0\rangle$ and $|1\rangle$ for each qubit. The walkers are initialized with a given momentum so that they evolve toward the right throughout the path that defines the circuit. Single-qubit gates can be implemented through single-particle scattering processes, e.g. through the use of appropriately designed graphs structures attached to the paths of the walker \cite{childs2013,asaka2023}. Then, two-qubit gates are implemented through two-particle scattering processes, such as the one described in Sec. \ref{sec:walker_theory}, in which the walkers are usually routed towards a common region of the circuit and interact with each other. This interaction induces a phase on the interacting states, so that, through the routing of the paths corresponding to the ket state $|1\rangle$ of each qubit, it generates a CP gate:
\begin{equation}\label{eq:cp-gate}
U_{\mathrm{CP}} = \mathrm{diag}(1,1,1,e^{i\theta})\, .
\end{equation}
This process is schematically illustrated in Fig. \ref{fig:fig1} through two representative implementations. The first one, in panel (a), was previously proposed in Ref. \cite{asaka2023}. There, the authors make use of roundabout gates acting on the internal state of the walkers (e.g. the $1/2$-spin of a fermion) to route two indistinguishable particles on the same path. An appropriate choice of walker momenta and interaction strength allows to implement the specified phase rotation [see Sec. \ref{sec:walker_theory}]. We also propose an alternative implementation based on distinguishable particles interacting on parallel chains, as illustrated in Fig. \ref{fig:fig1}(b). Here, the walkers are directed towards a common region by bending their paths and bringing them close together, however without any graph overlap. This eliminates the need of single-particle gates for the routing, thus reducing the fidelity loss in the case of localized wavepackets [see Sec. \ref{sec:finite_fidelity}]. After the interaction the two walkers are each moving along the other's path. Thus, they need to be switched back by exhanging the paths with each other.\newline 

Asaka et al. \cite{asaka2023} have shown that the setup in Fig. \ref{fig:fig1}(a) allows to accumulate a phase of $\pi/2$ (required for the implementation of a CNOT gate) with unitary fidelity, when considering definite momenta $q_1=-\pi/2$ and $q_2=\pi/2$ and on-site/nearest-neighbor interactions for bosons/fermions. However, practical implementations require a finite-size wavepacket composed of various components with different momenta, thus compromising the fidelity and modifying the phase of the gate. This consideration implies the need of a more flexible implementation, in which both the fidelity and phase can be finely tuned through additional degrees of freedom in the system, as to retrieve the ideal behavior for finite wavepackets. We see below that this is possible through the inclusion of additional interaction terms between the walkers. 

\subsection{Interacting walkers on parallel chains}\label{sec:walker_theory}

Before calculating gate fidelities, we start by briefly providing the theoretical description for the scattering of two interacting walkers. Specifically, we consider two interacting continuous-time QWs counter-propagating on 1D-chains consisting of $N$ nodes, $j,k\in[0,N-1]$. This corresponds to the situation within the blue regions of Figs. \ref{fig:fig1}(a) and (b). This system defines a Hilbert space with basis $\{|j,k\rangle : j,k \in [0, N-1]\}$, and is described by the following Hamiltonian:
\begin{equation}\label{eq:ham_tot}
    H = H_1\otimes\mathbb{1} + \mathbb{1}\otimes H_2 + H_{\rm int}\, .
\end{equation}
In the equation above, $H_i$ with $i=1,2$ are the single-particle tight-binding Hamiltonians:
\begin{equation}\label{eq:ham_single}
    H_i = J\sum_{j}(|j\rangle\langle j+1| + |j+1\rangle\langle j|)\, ,
\end{equation}
where $J$ is the hopping parameter between adjacent sites, and we set the on-site energy terms to zero without loss of generality. In the following, we set $J=1$, which is equivalent to rescaling the time parameter of the system to dimensionless units, $\tau \rightarrow t = J\tau/\hbar$. We remind that, for an infinite chain, the eigenvalues of the hamiltonian in Eq. \eqref{eq:ham_single} only depend on the particle wave vector $q_i\in[-\pi,\pi]$, and the eigenvectors are delocalized plane waves:
\begin{equation}
    E(q_i) = 2\cos(q_i)\, , \quad |\psi_{i,q_i}\rangle = \frac{1}{\sqrt{2\pi}}\sum_j e^{iq_ij}|j\rangle\, .
\end{equation}
At this point, we define the two-particle interacting Hamiltonian $H_{\rm int}$, which generally reads:
\begin{equation}
    H_{\rm int} = \sum_{j,k}w_{jk}|j,k\rangle\langle j,k|\, ,
\end{equation}
with $w_{j,k}$ the interaction terms. We consider a distance-dependent interaction, so that we can write:
\begin{equation}
    w_{jk} = \begin{cases}
        w_{|j-k|} & |j-k|\leq C \\
        0 & |j-k|> C
    \end{cases}\, ,
\end{equation}
where $C$ is the radius of the interaction.\newline

The model above defines a two-particle scattering problem with a finite interaction interval. In the ideal case of infinite chains, the scattering process can be understood by considering counter-propagating walkers with definite momenta $q_1\in (-\pi,0)$ and $q_2 \in (0,\pi)$. By defining the center of mass $M=(j+k)/2$ and the relative coordinate $r = j-k$, plus the total momentum $q=q_1+q_2$ and relative momentum $q_{\rm r}=(q_1-q_2)/2$, the two-particle wavefunction in the distinguishable case can be written in separable form:
\begin{equation}
    \langle M,r|\Psi(q,q_r)\rangle = e^{iMq}\langle r|\psi_q(q_r)\rangle\, ,
\end{equation}
where the state $|\psi_q(q_r)\rangle$ is an effective single-particle eigenstate of the relative coordinate Hamiltonian:
\begin{equation}\label{eq:ham_rel}
    H_r = \sum_r \left[2\cos(q/2)(|r\rangle\langle r+1| + h.c.) + w_r|r\rangle\langle r|\right]\, ,
\end{equation}
with eigenvalue $E(q,q_r)=4\cos(q/2)\cos(q_r)$ and shape:
\begin{align}\label{eq:state_rel}
    \langle r|\psi_q(q_r)\rangle = \begin{cases}
        e^{-iq_r r} + R_q(q_r)e^{iq_r r} & r\leq -C \\
        f_q(r;q_r) & |r|<C \\
        T_q(q_r)e^{-iq_r r} & r\geq C
    \end{cases}\, ,
\end{align}
In the equation above, $R_q(q_r)$ and $T_q(q_r)$ are the reflection and transmission coefficients, respectively. From now on, for simplicity of notation we will call $R=R_q(q_r)$, $T=T_q(q_r)$ and $f_r = f_q(r;q_r)$. The solutions of the scattering problem, i.e. the values of $T$ and $R$, can be found analytically by solving the eigenvalue equation of the Hamiltonian in Eq.  \eqref{eq:ham_rel} with the ansatz in \eqref{eq:state_rel} [see Appendix \ref{app:scattering} for explicit solutions for $C\leq 2$].\newline

\subsubsection{Indistinguishable particles}\label{sec:indist_ptc}
Up to now we have described the situation for distinguishable particles, i.e. walkers evolving on different parallel chains, for which the Hamiltonian in Eq. \eqref{eq:ham_tot} acts on the two-particle Hilbert space. To pass to bosonic/fermionic walkers evolving on the same chain, one must impose symmetry/antisymmetry relations under the exchange of the particles. This is possible through the use of the following projection operators:
\begin{equation}
    P_{\rm S} = (\mathbb{1}+P_{12})\, , \quad P_{\rm A}=(\mathbb{1}-P_{12})\, ,
\end{equation}
where $P_{12}$ is the swap operator:
\begin{equation}
    P_{12} |j,k\rangle = |k,j\rangle \, .
\end{equation}
Specifically, we can symmetrize/antisymmetrize the expression of the ansatz in Eq. \eqref{eq:state_rel}, obtaining:
\begin{align}\label{eq:state_rel_dist}
    \langle r|\psi_q^\pm(q_r)\rangle =& \frac{1}{\sqrt{2}}(\langle r|\psi_q(q_r)\rangle \pm \langle -r|\psi_q(q_r)\rangle) \nonumber \\
    =&\begin{cases}
        e^{-iq_r r} \pm S^\pm e^{iq_r r} & r\leq -C \\
        f_r\pm f_{-r} & |r|<C \\
        S^{\pm}e^{-iq_r r}\pm e^{iq_rr} & r\geq C
    \end{cases}\, ,
\end{align}
where we defined:
\begin{equation}\label{eq:S+-}
    S^\pm\equiv T\pm R =e^{i\theta^\pm}\, .
\end{equation}
The expression above comes from the fact that $|T+R|=1$ due to unitarity. Consequently, in the indistinguishable case, the scattered wave function simply possesses an induced phase $\theta^\pm$.

\subsection{Gaussian wavepackets}\label{sec:wavepacket_theory}

In this paper we aim at simulating the CP gate for finite wavepackets, as to accurately and quantitively account for finite-size effects on the fidelity, and how fidelity loss can be overcome by tuning the interaction potential between walkers. In order to do so, we choose a Gaussian wavepacket, which 
which most closely maintains its shape in time during the evolution. At a reference initial time $t=0$, we define the state of each walker as:
\begin{equation}\label{eq:gaussian}
    \langle j|\psi_i(0)\rangle = \frac{1}{(2\pi\tilde{\sigma}^2)^{1/4}} e^{-\frac{(j-j_i)^2}{4\tilde{\sigma}^2}}e^{iq_ij}\, ,
\end{equation}
where $j_i$ is the initial central position of the walker, $q_i$ its central momentum, and $\tilde{\sigma}$ its standard deviation. The shape of each wavepacket in momentum space can be found through a Fourier expansion of the basis states, and is also a Gaussian:
\begin{equation}\label{eq:psi_momentum}
    \langle k|\psi_i(0)\rangle = \frac{1}{(2\pi\tilde{\sigma}_k^2)^{1/4}} e^{-\frac{(k-q_i)^2}{4\tilde{\sigma}_k^2}}e^{ikj_i}\, ,
\end{equation}
which is centered around $q_i$ and with standard deviation $\tilde{\sigma}_k=1/2\tilde{\sigma}$. Once we have defined the initial state of each walker, its evolution in time can be evaluated through the Schr\"odinger equation. For a single particle:
\begin{equation}
    |\psi_i(t)\rangle = e^{-iH_it}|\psi_i(0)\rangle\, ,
\end{equation}
where $H_i$ is the single-particle Hamiltonian from Eq. \eqref{eq:ham_single}, and we used the dimensionless definition of time $t$ previously introduced in Sec. \ref{sec:walker_theory}. For the two-particle state, which we initialize as a product state:
\begin{equation}
    |\Psi(0)\rangle = |\psi_1(0)\rangle\otimes|\psi_2(0)\rangle\, ,
\end{equation}
the evolution is instead described by the total Hamiltonian from Eq. \eqref{eq:ham_tot}. Considering the Gaussian shapes of the wavepackets from Eq. \eqref{eq:gaussian}, the two-particle distinguishable wavefunction can also be written in terms of the center-of-mass and relative coordinates:
\begin{align}\label{eq:two_p_wf}
    \langle M,r|\Psi(0)\rangle =&  \frac{1}{(2\pi\sigma_c^2)^{1/4}}e^{-\frac{-(M-M_0)^2}{4\sigma_c^2}}e^{iqM} \nonumber\\ &\quad\times\frac{1}{(2\pi\sigma^2)^{1/4}}e^{-\frac{(r-r_0)^2}{4\sigma^2}}e^{-iq_rr} \, ,
\end{align}
where $\sigma_c = \tilde{\sigma}/\sqrt{2}$ and $\sigma = \sqrt{2}\tilde{\sigma}$ are respectively the standard deviations for the center-of-mass and relative coordinates. In the following, we will mostly use $\sigma$ as the reference quantity to describe the system of the interacting finite wavepackets, as this is the main quantity that influences the fidelity.\newline

Each component of the wavepacket with momentum $k_i$ possesses a different velocity due to the shape of the single-particle band structure:
\begin{equation}
    v(k_i) = \frac{\partial E(k_i)}{\partial k_i}=-2\sin(k_i)\, .
\end{equation}
Because of this, even in the absence of interactions, the wavepacket inevitably loses its shape in time depending (mainly) on the second derivative of the band. Indeed, the standard deviation after time $t$ can be approximated for a narrow momentum distribution ($\tilde{\sigma}_k\ll1$) as:
\begin{equation}\label{eq:sigma_t}
    \tilde{\sigma}_t^2 \simeq \tilde{\sigma}^2(1+\eta^2)\, ,
\end{equation}
where $\eta = t/2\tilde{\sigma}^2m^*$ and $m^* = \left( \partial_{k_i}^2E(k_i)\big|_{q_i} \right)^{-1}$. Thus, the standard deviation of each wavepacket steadily increases in time, by a factor that is inversely proportional to itself and to the second derivative of the band structure at $q_i$. This implies that the best choice for the central momenta of the wavepackets is that for which the second derivative of the single-particle band is zero, i.e. $q_i=\pm \pi/2$. In the following, we will set a value of $\sigma=6$ ($\tilde{\sigma} = 3\sqrt{2}$). This value is small enough to permit relatively small circuit size, while being large enough so that the approximation in Eq. \eqref{eq:sigma_t} can be considered valid.

\begin{figure}[b!]
    \centering
    \includegraphics[width=\linewidth]{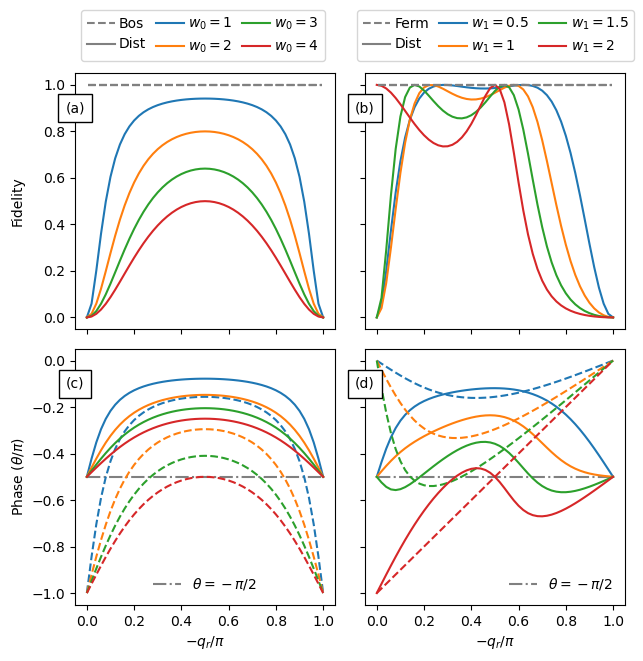}
    \caption{Ideal CP gate performance for plane-wave scattering with $q_1=-q_2=q_r$. (a) Fidelity and (c) induced phase for distinguishable particles (continuous lines) interacting through an on-site potential ($C=0$), and comparison with the corresponding bosonic implementation (dashed lines). (b) Fidelity and (d) induced phase for distinguishable particles (continuous lines) with first-neighbor interactions ($C=1$, $w_0=w_1$), and comparison with the corresponding fermionic implementation.}
    \label{fig:fid_ideal_C01}
\end{figure}

\section{Fidelity calculations}\label{sec:fidelity_results}

\subsection{Ideal CP gate fidelities}\label{sec:ideal_fidelity}
We start our discussion on the numerical values of the fidelity of a CP gate by showing the ideal values for plane-wave particles described by definite momenta $q_1\in(-\pi,0)$ and $q_2\in(0,\pi)$. Generally speaking, we can define the fidelity of a gate by comparing its output state with the expected ideal output:
\begin{equation}\label{eq:fid_dist}
    F = |\langle\Psi_{\rm ideal}(q,q_r)|\Psi_{\rm out}(q,q_r)\rangle |^2\, ,
\end{equation}
where the ideal output of a CP gate possesses a relative phase with respect to the free-state at time $t$:
\begin{equation}
    |\Psi_{\rm ideal}(q,q_r)\rangle = e^{i\theta_0}|\Psi_{\rm free}(q,q_r)\rangle\, ,
\end{equation}
with $\theta_0$ being the induced phase, which is the argument of either the scattering matrix or transmission coefficient, respectively in the indistinguishable and distinguishable case. For indistinguishable particles with definite momenta, using the property in Eq. \eqref{eq:S+-}:
\begin{equation}
    |\Psi_{\rm out}^\pm(q,q_r)\rangle = S^\pm|\Psi_{\rm free}(q,q_r)\rangle\, ,
\end{equation}
\begin{equation}
    F_{\rm \pm} = |S^\pm|^2 = 1\, ,
\end{equation}
so that the fidelity is identically one. Instead, the real output state for distinguishable particles is described by the transmitted part of the interacting wavefunction, so that:
\begin{equation}
    |\Psi_{\rm out}^{\rm dist}(q,q_r)\rangle = T|\Psi_{\rm free}(q,q_r)\rangle\, ,
\end{equation}
\begin{equation}
    F_{\rm dist} = |T|^2 \leq 1\, .
\end{equation}
Thus, the fidelity for distinguishable particles explicitly depends on the transmission coefficient, and can be equal to 1 only when $|T|=1$, that is, when the particles are perfectly transmitted.\newline

In Fig. \ref{fig:fid_ideal_C01} we show the fidelity from Eq. \eqref{eq:fid_dist} for on-site and first-neighbor interactions. In both situations, we compare the results from the distinguishable case with those from the indistinguishable case (bosons for $C=0$ and fermions for $C=1$), which were also previously reported in Refs. \cite{childs2013,asaka2023}. For all cases, we set $q_1 = -q_2 = q_r$, and only show the results for $q_r\in[-\pi,0]$, though the situation is symmetric for $q_r\in[0,\pi]$. As we mentioned above, both bosons and fermions always show unitary fidelity. Specifically, at $q_r=-\pi/2$, bosons possess an induced phase of $\theta=-\pi/2$ for $w_0=4$, while fermions possess the same phase for $w_1=2$. For distinguishable particles instead, the situation is radically different. As we see from Fig. \ref{fig:fid_ideal_C01}(a), the fidelity is indeed always strictly lower than 1 for $C=0$, showing a fixed maximum at $q_r=-\pi/2$. In this case, a larger interaction progressively induces particle reflection, so that the fidelity decreases. At the same time, Fig. \ref{fig:fid_ideal_C01}(c) shows that the phase induced on the transmitted particles increases with the interaction, but reaches the ideal value of $\theta=-\pi/2$ only at $q_r=0$ and $q_r=\pi$, or for asymptotic values of $w_0\rightarrow \infty$, all cases for which $F$ is identically zero. All of these results imply that distinguishable particles in the on-site interaction regime are practically unsuitable for a CP gate implementation.\newline

However, even the introduction of a single additional interaction term changes the situation. Indeed, in Fig. \ref{fig:fid_ideal_C01}(b) we considered nearest-neighbor interaction ($C=1$), initially setting for simplicity $w_1=w_0$. As it is apparent, this time the fidelity for distinguishable particles reaches 1 for multiple values of $q_r$. Studying the solutions of $T$ for $C=1$ from the eigenvalue equation of the Hamiltonian in Eq. \eqref{eq:ham_rel}, one indeed finds that with the restriction that $w_0=w_1$ there exist four solutions for which $|T|=1${, which we call $\pm q_{r,1}$ and $\pm q_{r,2}$, with $q_{r,1}<q_{r,2}$ [their explicit expressions are reported in Appendix \ref{app:unitary_fidelity}]. As the interaction increases, all solutions shift towards the origin, and show increasing curvature of the fidelity, which consequently decays faster with respect to its maximum. The solutions $\pm q_{r,1}$ exist only for $w_0=w_1\in[-6, 2]$, while $\pm q_{r,2}$ exist only for $w_0=w_1\in[-2, 6]$, otherwise the fidelity is strictly smaller than 1 for all $q_r$}. Importantly, for $q_r=-\pi/2$ and $w_0=w_1=2$, the fidelity is 1 and the induced phase on the particles is exactly $-\pi/2$ [see Fig. \ref{fig:fid_ideal_C01}(d)], as it was in the fermionic case. Consequently, the inclusion of additional interaction terms opens the way for the implementation of CP gates with distinguishable particles, reaching the same performances of indistinguishable particles, at least in the ideal case of plane waves with definite momenta. Conceptually, this behavior can be explained in terms of the interaction being more distributed in space with respect to the on-site case. Indeed, it has been shown that interacting quantum walks with accurately-tailored long-range potentials are able to maintain correlation effects and entanglement-generation capabilities on par with on-site interactions, while reducing the detrimental effects on transport \cite{forghieri2026}.\newline

\begin{figure}[t]
    \centering
    \includegraphics[width=\linewidth]{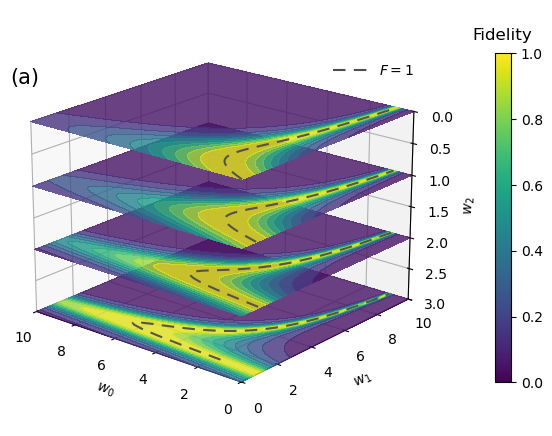}
    \includegraphics[width=\linewidth]{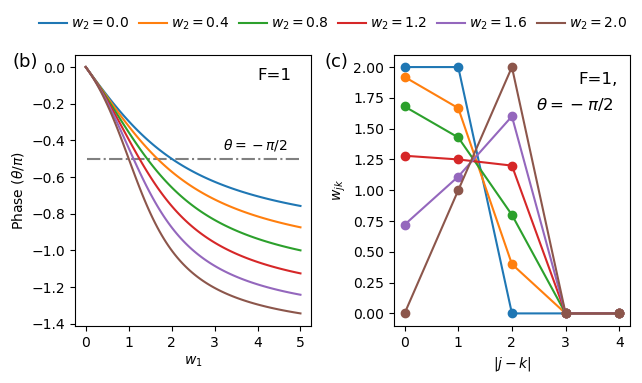}
    \caption{Ideal CP gate performance for plane-wave scattering for distinguishable particles with interaction radius $C=2$ and momenta $q_1=-q_2=-\pi/2$. (a) Fidelity as a function of the interaction terms $w_0$, $w_1$, and $w_2$. The dashed grey curves indicate the loci of points for which $F=1$ for each $w_2$-plane. (b) Conditional phase $\theta$ accumulated along the unit-fidelity solutions. (c) Corresponding interaction profiles yielding $\theta=-\pi/2$, required for the implementation of a CNOT gate.}
    \label{fig:fid_ideal_C2}
\end{figure}

We generalize the previous discussion in Fig. \ref{fig:fid_ideal_C2}, where we show the ideal behavior of a CP gate for distinguishable particles for $C=2$, i.e. with the interaction extended to second nearest neighbors. In this case, we set $q_1=-q_2=-\pi/2$. This corresponds to the most favorable setting both in terms of the previously shown results and of the group velocity introduced in Sec. \ref{sec:wavepacket_theory}, which will later come into play when considering finite wavepackets. In Fig. \ref{fig:fid_ideal_C2}(a) we show the fidelity of the CP gate as a function of different interaction terms $w_0$, $w_1$ and $w_2$ (restricted to positive values for easier representation). Notice that, whatever the value of the second-neighbor term may be, there always exists a locus of points in the $w_0-w_1$ plane that produces exactly unit fidelity, although shifted to larger $w_1$ and progressively presenting more pronounced peaks. The derivation of the condition for unitary fidelity is explicitly derived in Appendix \ref{app:unitary_fidelity}. In Fig. \ref{fig:fid_ideal_C2}(b) we show the phase induced on the particles along these loci of points. As we can see, any phase in the interval $[-\pi,0]$ can be easily achieved by appropriately tuning the three parameters of the interactions.Coherently, any positive phase in the interval $[0,\pi]$ is also achievable by replacing $w_r\rightarrow -w_r$. This ultimately shows that it is possible to implement any arbitrary controlled-phase rotation as in Eq. \eqref{eq:cp-gate} with distinguishable particles, maintaining the same fidelity as the indistinguishable case. Finally, in Fig. \ref{fig:fid_ideal_C2}(c), we illustrate the shape of the potentials producing an exact phase of $\theta=-\pi/2$ [see derivation in Appendix \ref{app:unitary_fidelity}], which is the one required for the implementation of the CNOT gate. Notice that it is not necessary for the potential to be monotonically decreasing with respect to the distance. On the contrary, for $w_2>\sqrt{5}-1\simeq 1.24$, the shape of the potential is peaked at $r=2$. \newline

In the next section, we extend this ideal plane-wave description to realistic finite wavepackets, where dispersion and finite-size effects must be taken into account.

\begin{figure}[b!]
    \centering
    \includegraphics[width=\linewidth]{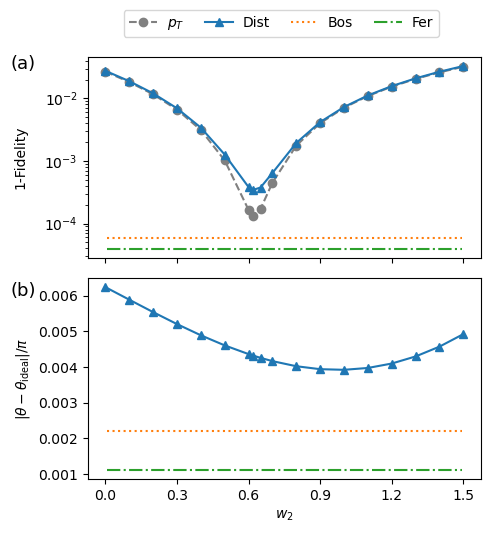}
    \caption{Finite-wavepacket effects on the CP gate fidelity for Gaussian wavepackets with second-neighbor interactions and central momenta $q_1=-q_2=-\pi/2$. For each value of $w_2$, the interaction parameters $w_0$ and $w_1$ are chosen as to reproduce the ideal unitary fidelity for an induced phase of $\theta=-\pi/2$, as in Fig. \ref{fig:fid_ideal_C2}(c). (a) Fidelity loss $1-F$ of distinguishable particles. The panel shows a comparison with the transmission loss $1-p_T$ of the gate, and with bosonic (on-site, $w_0=4$) and fermionic (first-neighbor, $w_1 = 2$) implementations. (b) Relative deviation of the accumulated phase from the ideal value $\theta_{\rm ideal}=-\pi/2$.}
    \label{fig:fid_sigma6}
\end{figure}

\subsection{Finite-site effects on fidelity}\label{sec:finite_fidelity}
For an accurate description of the CP gate behavior in the presence of finite wavepackets, we simulated the interaction regions from Fig. \ref{fig:fig1} as chains consisting of $N=101$ nodes, setting the relative-coordinate standard deviation to $\sigma=6$ (so that each wavepacket has $\tilde{\sigma}=3\sqrt{2}$). By initializing the first wavepacket at the central position $j_1 = 30$ and the second one at $j_2=70$, these conditions allow to neglect border effects and particle overlaps at $t=0$. Numerically, we constructed each wavepacket in the interval $j\in[j_i-3\sigma,j_i+3\sigma]$. We then set the particles' central momenta to $q_1=-q_2=-\pi/2$. The numerical values of the fidelities reported below were then evaluated by evolving the two-particle wavefunction at time $t=20$.\newline

Since the two-particle wavefunction from Eq. \eqref{eq:two_p_wf} is separable in the center-of-mass and relative coordinates, and the scattering process only affects the relative-coordinate wavefunction, we can focus on the latter to define the CP gate fidelity for finite wavepackets. In this case, the ideal output of the gate can be written by taking account of its effective length $\theta_1$:
\begin{equation}
    |\psi_{{\rm ideal},r}\rangle = e^{i\theta_0}e^{i\theta_1\hat{k}_r}|\psi_{{\rm in},r}\rangle\, ,
\end{equation}
where $\theta_1$ is defined as the first derivative of the phase of $T$:
\begin{equation}
    \theta_1 = \frac{\partial \theta_q(k_r)}{\partial k_r}\bigg|_{-q_r}\, .
\end{equation}
This quantity causes a momentum-dependent delay in the wavepacket. Such a contribution can be compensated through suitable path-length adjustments and therefore does not represent an intrinsic gate error. That said, it needs to be carefully taken into account when designing finite circuits working with wavepackets, as synchronization is crucial. Because of this, when we evaluate the fidelity numerically, we need to compare the final state of the interacting particles with a shifted free-evolving state, with particles initialized respectively at $j_1+\theta_1/2$ and $j_2-\theta_1/2$.\newline

\begin{figure}[t]
    \centering
    \includegraphics[width=\linewidth]{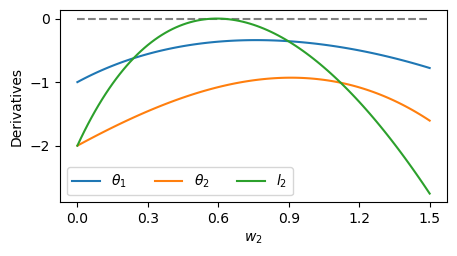}
    \caption{Quantities governing the Gaussian-wavepacket fidelity of the distinguishable-particle CP gate under the same conditions as Fig. \ref{fig:fid_sigma6}. Shown are the effective interaction length $\theta_1$ (blue), the phase-curvature term $\theta_2$ (orange), and the logarithmic-magnitude curvature $l_2$ (green), the latter two entering the approximated fidelity expression in Eq. \eqref{eq:fid_dist_wp}. All quantities are evaluated at $-q_r=\pi/2$, at which the logarithmiq-magnitude first derivative $l_1$ is identically zero.}
    \label{fig:der_dist}
\end{figure}

In Fig. \ref{fig:fid_sigma6}(a) we show the numerical fidelities obtained from our simulations. For these trials, we considered a general second-neighbor interaction for the distinguishable case, and set the interaction terms corresponding to the loci of points from Fig. \ref{fig:fid_ideal_C2} that provide unitary fidelity and induced phase $\theta=-\pi/2$ for plane waves. We also report the reference values of the fidelities bor bosonic and fermionic wavepackets, respectively setting an on-site interaction with $w_0=4$ and a nearest-neighbor interaction with $w_1=2$. As we see from the picture, a careful choice of the interaction potential for distinguishable particles helps reduce the fidelity loss with respect to the plane-wave case. Specifically, the maximum fidelity is obtained for $w_2\simeq 0.6$, for which $1-F\simeq 3.79$e-4. Instead, Fig. \ref{fig:fid_sigma6}(b) shows that the discrepancy with respect to the ideal phase of $-\pi/2$ only slightly changes. In order to better understand the mechanisms behind fidelity loss, we can study the analytical expression of the fidelity or Gaussian wavepackets. Specifically, by introducing the spectral decomposition of the identity in the definition of the fidelity and focusing on the relative-coordinate part of the wavefunction, we get:
\begin{align}\label{eq:fidelity_wp_dist}
    F_{\rm dist} =&\, |\langle\psi_{{\rm ideal},r}|\psi_{{\rm out},r}\rangle |^2\, \nonumber \\
    =&\, \left|\int_{-\infty}^{\infty}e^{-i(\theta_0+\theta_1k_r)}\langle\psi_{{\rm in},r}|k_r\rangle T_q(k_r)\langle k_r|\psi_{{\rm in},r}\rangle\, dk_r\,\right|^2 \nonumber \\
    =&\, \left|\int_{-\infty}^{\infty}e^{-i\theta_1k_r}P(k_r)\,T_q(k_r)\, dk_r\,\right|^2\, ,
\end{align}
where $P(k_r) = |\langle k_r|\psi_{{\rm in},r}\rangle|^2$ is a probability distribution based on the normalized relative-coordinate wavefunction in momentum space. To complete this calculation, one can expand the transmission coefficient by taking account of both its modulus and phase changing as a function of $q_r$. By doing so, one obtains an approximated expression for the fidelity, valid for $\tilde{\sigma}_k\ll 1$ [see derivation in Appendix \ref{app:finite_fidelity}]:
\begin{align}\label{eq:fid_dist_wp}
    F_{\rm dist}\simeq& |T_q(-q_r)|^2\exp[(l_1^2+2l_2)\sigma_k^2]\exp(-2\theta_2^2\sigma_k^4) \nonumber \\
    =& F_{\rm magn}(T,l_1,l_2) F_{\rm phase}(\theta_2)
\end{align}
where $\sigma_k=1/2\sigma$ is the relative-coordinate standard deviation in momentum space; $l_1/l_2$ are the first/second derivatives of the logarithmic magnitude of $T$, and $\theta_2$ is the second derivative of its phase, evaluated at momentum $-q_r$. In Fig. \ref{fig:der_dist} we report the values of $\theta_2$ and $l_2$ defined above under the same settings of Fig. \ref{fig:fid_sigma6}, along with the value of the effective length $\theta_1$, for completeness. Notice that, in the considered situation, $l_1\equiv0$ and $l_2\leq 0$, since we set a relative momentum for which the transition amplitude is maximum. As we can see, $l_2$ shows a maximum right at $w_2=0.6$, thus explaining the minimum in Fig. \ref{fig:fid_sigma6}(a).\newline

A similar procedure to the one described above can also be used to evaluate the transmission probability $p_T$ [see Appendix \ref{app:finite_transmission}]:
\begin{align}\label{eq:pt_dist_wp}
    p_T\simeq F_{\rm magn}\exp(l_1^2\sigma_k^2)\leq 1\, .
\end{align}
By comparing Eqs. \eqref{eq:fid_dist_wp} and \eqref{eq:pt_dist_wp}, one actually finds that the gate fidelity is always upper limited by $p_T$ and, even when $l_1=0$ as in this case, $F$ will be slightly smaller due to the phase contribution to the fidelity loss. This is perfectly consistent with the numerical results of $p_T$ shown in Fig. \ref{fig:fid_sigma6}(a), and found through a umerical integral of the transmitted wavefunction after the interaction. As for the indistinguishable case, one can write an equivalent formula to Eq. \eqref{eq:fidelity_wp_dist} by replacing $T$ with $S^\pm$ and $\theta_1$ with $\theta_1^\pm$:
\begin{equation}
    F_\pm = \left|\int_{-\infty}^{\infty}e^{-i\theta_1^\pm k_r}P(k_r)\,S^{\pm}_q(k_r)\, dk_r\,\right|^2\, .
\end{equation}
In this case, the calculation is the same, except for the fact that the magnitude of $S^\pm$ is fixed to one and $l_1/l_2=0$. Thus, we get an approximated fidelity of:
\begin{align}\label{eq:fidelity_indist}
    F_\pm \simeq \exp(-2\theta_2^2\sigma_k^4) = F_{\rm phase}(\theta_2)\, .
\end{align}
Because the fidelity in this case has no magnitude contribution, and instead depends on $\sigma_k^4$ at its higher order, we see in Fig. \ref{fig:fid_sigma6} that the fidelity loss is lower of almost an order of magnitude with respect to the minimum one from the distinguishable case. The phase discrepancy is also lower [see Fig. \ref{fig:fid_sigma6}(b)], although only of a factor of $\sim2$ and $\sim 4$ respectively.\newline

Overall, from the previous formulas we see that the fidelity loss in the case of finite wavepackets has several contributions. First of all, the variation in the scattering phase acquired by each wavepacket component, and specifically its curvature $\theta_2$, causes 
wavepacket \textit{broadening/compression}. The other contributions, caused by the change in transmission amplitude, are only present in the case of distinguishable particles, and are much more impacting on the fidelity, being dependent on $\sigma_k^2$ rather than on $\sigma_k^4$. These are generally able to increase the fidelity, either through $l_1\neq 0$ or $l_2>0$. However, through considerations based on the transmission probability $p_T$, one sees that the only case in which the wavepacket fidelity can reach 1 is for $l_1=0$. If $T(q_r)=1$, this condition is satisfied: however, one will necessarily have $l_2\leq 1$, so that in this case the magnitude contributions will also almost inevitably cause a loss of fidelity.\newline

\begin{table}[t]
\begin{tabular}{l|c|c|c|c|c|}
            & Log. Mag.  & \multicolumn{2}{c|}{Phase} & \multirow{2}{*}{$1-F_{\rm th}$} & \multirow{2}{*}{$1-F_{\rm num}$} \\ \cline{2-4}
            & 2nd deriv. & 1st deriv. & 2nd deriv.    &  & \\
\hline
Bos.      & 0          & 0           & -0.5         & 2.41e-5     & 5.53e-5 \\
Ferm.    & 0          & 1           & 0            & 0                & 3.69e-5 \\
R$^{(1)}$  & -0.25      & 3           & 0            & 6.92e-3     & 3.47 e-3 \\
Dist.    & -1.37e-4   & -0.363      & -1.08        & 1.14e-4     & 3.79e-4
\end{tabular}
\caption{Relevant derivatives of the scattering/transmission amplitudes for the different CP gate implementations considered in this work, evaluated at $q_1=-q_2=-\pi/2$ and at $-q_r=\pi/2$. Beside are reported the corresponding theoretical ($1-F_{\mathrm{th}}$) and numerical ($1-F_{\mathrm{num}}$) fidelity losses for Gaussian wavepackets.}
\label{table:t1}
\end{table}

\subsubsection{Fidelity of the roundabout-assisted CP gate}\label{sec:roundabout_fidelity}
The results reported above would lead one to believe that the distinguishable case is inevitably inferior to the bosonic or fermionic implementation of the CP gate. However, sontrarily to the distinguishable case, the setup in Fig. \ref{fig:fig1}(a) for indistinguishable particles also requires two single-qubit gates acting on each walker in order to implement a CP gate. Therefore, when evaluating the fidelity of the CP gate for the indistinguishable case, one must also account for finite-size effects on the roundabout gate from Ref. \cite{asaka2023}, whose fidelity is [see derivation in Appendix \ref{app:roundabout_fidelity}]:
\begin{align}\label{eq:roundabout_fid}
    F_{\rm R}\simeq& |S_{\rm R}(q_i)|^2\exp[2(m_1^2+2m_2)\sigma_k^2]\exp(-8\phi_2^2\sigma_k^4) \nonumber \\
    =& F_{\rm R,magn}(S_{\rm R},m_1,m_2)F_{\rm R,phase}(\phi_2)\, ,
\end{align}
where $m_1/m_2$ are the first/second derivatives of the logarithmic magnitude of the single-particle scattering matrix element $S_{\rm R}(k)$, and $\phi_2$ is the second derivative of its phase, evaluated at the single-particle central momentum $q_i$. Notice that the fidelity in Eq. \eqref{eq:roundabout_fid} is evaluated with respect to the relative two-particle standard deviation, $\sigma_k = \tilde{\sigma}_k/\sqrt{2}$. Both the numerical and theoretical values of this fidelity are reported in Table \ref{table:t1} for $|q_i|=\pi/2$ and compared to the fidelities from Fig. \ref{fig:fid_sigma6}(a). As one can see, the fidelity loss of the roundabout gate in the presence of a Gaussian wavepacket is higher of approximately an order of magnitude with respect to the one for the scattering of distinguishable particles. This effect is mainly due to the value of $m_2$, which differs of about three orders of magnitude with respect to $l_2$, and thus produces a larger fidelity loss.\newline

Putting everything together, one can also find the total fidelity from the setup in Fig. \ref{fig:fig1}(a) by combining the scattering coefficients from both the roundabout gates for both particles and the two-particle scattering for indistinguishable particles:
\begin{equation}\label{eq:tot_scattering_indist}
S_{\rm tot}(k_1,k_2)
=
S_{\rm R}^{(2)}(k_1)\,
S_{\rm R}^{(2)}(k_2)\,
S^{\pm}(k,k_r)\,
S_{\rm R}^{(1)}(k_1)\,
S_{\rm R}^{(1)}(k_2),
\end{equation}
where $S_R^{(1)}$ and $S_R^{(2)}$ denote the first and second roundabout gates that each particle traverses, and $S^{\pm}$ is the two-particle scattering coefficient. The total fidelity of the setup in Fig.~\ref{fig:fig1}(a) cannot generally be obtained as the product of the individual gate fidelities, since the roundabout and two-particle scattering coefficients contribute jointly to the final scattering amplitude. The roundabout-induced fidelity loss and the interaction-induced fidelity loss can be analysed separately, but the fidelity of the complete architecture is intrinsically a correlated two-particle quantity and depends on mixed phase derivatives generated by the interaction region [see Appendix \ref{app:combined_fidelity} for more details]. That said, in the case of $|\theta_2|\sigma_k^2\ll 1$, one can approximate the total fidelity as the product of the individual roundabout and 
interaction fidelities (assuming all four roundabout gates have identical fidelity):
\begin{equation}\label{eq:ftot_prod}
    F_{\rm tot} \simeq F_\pm(F_{\rm R})^4\, ,
\end{equation}
which, in this case, would give an approximate fidelity loss of $1-F_{\rm tot, th}\simeq 2.74\, 10^{-2}$, both for bosons and fermions. \newline

Overall, the possibility of implementing a CP gate by means of distinguishable particles allows to reduce the number of elements necessary to build the architecture from Fig. \ref{fig:fig1}(b). Thus, even though the scattering process alone is less performing than the one for indistinguishable particles, the overall distinguishable architecture significantly reduces spurious fidelity losses from single-particle gates. 

\section{Conclusions}\label{sec:conclusions}
{\color{black} We have analyzed the implementation of controlled-phase gates in dual-rail quantum walk architectures, focusing on the role of interaction range and particle statistics in determining gate fidelity. Our main result is that extending the interaction beyond nearest neighbors enables the engineering of arbitrary controlled-phase rotations with distinguishable particles while maintaining unit fidelity for plane-wave scattering. Specifically, for second-neighbor interactions, we have identified interaction parameters for which the transmission coefficient goes to unit, and shown that the induced phase can be continuously tuned across the whole interval $[-\pi,\pi]$ by appropriately choosing the interaction strengths. This provides a systematic approach to implementing any CP gate without relying on particle indistinguishability.

For finite Gaussian wave-packets, we derived analytical expressions for the gate fidelity in terms of the transmission coefficient's magnitude and phase derivatives. The fidelity separates into magnitude and phase contributions. This hierarchy explains why the distinguishable case, which suffers from both magnitude and phase contributions, exhibits larger fidelity losses than the indistinguishable case, where only phase variations contribute. Nevertheless, for optimized interaction parameters, we found that fidelity losses for distinguishable particles is only about one order of magnitude larger than the bosonic/fermionic values.

Crucially, when accounting for the full experimental architecture, the roundabout gates required for routing indistinguishable particles introduce additional fidelity losses nearly two orders of magnitude larger than the two-particle scattering losses. For the complete CP gate requiring four roundabout gates, the total fidelity loss is dominated by single-particle gate imperfections. The distinguishable-particle architecture avoids these additional elements, making its overall performance competitive despite the slightly lower intrinsic scattering fidelity.



Our analysis and results also establish a foundation for several natural extensions. For instance, the analytical treatment developed for second-neighbour interactions can be directly ported to longer-range couplings, where it is expected to offer additional flexibility for fidelity optimization. Likewise, the inclusion of disorder and imperfections, while outside the present scope, can be systematically addressed within our formalism, paving the way for potential experimental validation. Finally, the modularity of our architecture supports generalization to multi-qubit gates, providing tools for assessing the associated error propagation in future work.

In conclusion, multi-neighbor interacting quantum walks provide a robust platform for implementing CP gates with finite wavepackets. The trade-off between intrinsic scattering fidelity and architectural complexity favors the distinguishable-particle approach when single-particle gate errors are taken into account. Our results establish quantitative benchmarks for experimental realizations and offer a theoretical framework for optimizing quantum gate operations in interacting many-particle systems.}

\section{Appendix}
\subsection{Explicit solutions for scattering with multi-neighbor interactions}\label{app:scattering}

In order to find the solutions of the scattering problem introduced in Sec. \ref{sec:walker_theory}, i.e. the coefficients $R$ and $T$, one must solve the eigenvalue equation for the Hamiltonian in Eq. \eqref{eq:ham_rel} with the ansatz in Eq. \eqref{eq:state_rel}. For the calculations, it is useful to define the following quantities:
\begin{align}
    \varepsilon_r =& \, \frac{[w_r-E(q,q_r)]}{2\cos(q/2)}\, ; \\
    \eta_r =& \, e^{-iq_r}+\varepsilon_r\, ; \\
    \xi_r =& \, e^{-iq_rr}\, .
\end{align}
For $C=0$ (Bose-Hubbard model), the solutions are \cite{childs2013}:
\begin{align}
    T =& \, -\frac{2i\sin(q_r)}{\eta_0 + \xi_1}\, ; \\
    R =& \, T-1\, ; \label{eq:continuity}
\end{align}
Note that Eq. \eqref{eq:continuity} is due to the continuity of the function in $r=0$. Generalizing to $C>0$ is equivalent to solving a system of $2C+1$ equations. Considering the matrix form $A\mathbf{x}=\mathbf{b}$, with $\mathbf{x}=(R,f_{-C+1},\dots,f_{C-1},T)^{\rm T}$ the vector of unknown variables, we have:
\begin{align}
    &A = \begin{pmatrix}
        \eta_C\xi_C & 1 & 0 & 0 & & & \\
        \xi_C & \varepsilon_{C-1} & 1 & 0 & & & \\
        0 & 1 & \varepsilon_{C-2} & 1 & & & \\
        & & \ddots & \ddots & \ddots & &  \\
        & & & 1 & \varepsilon_{C-2} & 1 & 0 \\
        & & & 0 & 1 & \varepsilon_{C-1} & \xi_C \\
        & & & 0 & 0 & 1 & \eta_C\xi_C
    \end{pmatrix}\, ,
\end{align}
\begin{align}
    \mathbf{b} = (-\eta_C^*\xi_C^*,-\xi_C^*,0\dots,0,0,0)^T\, .
\end{align}

Here, we explicitly report the analytical solutions for $C=1,2$, which are still relatively simple to find analytically. For $C=1$ (nearest-neighbor interaction):
\begin{align}\label{eq:t1}
    T =& \, -\frac{2i\xi_2^*\sin(q_r)}{\eta_1(\varepsilon_0\eta_1-2)}\, ; \\
    R =& \, (\varepsilon_0\eta_1-1)T -\xi_2^*\, ;
\end{align}
and for $C=2$:
\begin{align}\label{eq:t_c2}
    T =&\, -\frac{2i\xi_4^*\sin(q_r)}{[\varepsilon_0(\varepsilon_1\eta_2-1)-2\eta_2](\varepsilon_1\eta_2-1)}\, ; \\
    R =&\, [(\varepsilon_1\varepsilon_0-1) (\varepsilon_1\eta_2-1)-\varepsilon_1\eta_2]T-\xi_4^*\, .
\end{align}\\

\subsubsection{Indistinguishable particles}\label{app:scattering_indist}
As mentioned in Sec. \ref{sec:indist_ptc}, the case of indistinguishable particles can be straightforwardly derived from the results of the distinguishable case, by using the expressions of $R$ and $T$ written above. Thus, for $C=0$ (only bosons):
\begin{equation}
    S^{+} = -\frac{\eta_0^*+\xi_1^*}{\eta_0+\xi_1}\, ;
\end{equation}
for $C=1$:
\begin{equation}
    S^+ = -\xi_2^*\frac{(\varepsilon_0\eta_1^*-2)}{\varepsilon_0\eta_1-2}\, , \quad S^- = \xi_2^*\frac{\eta_1^*}{\eta_1}\, ;
\end{equation}
and for $C=2$:
\begin{equation}
    S^+ = -\xi_4^*\frac{\varepsilon_0(\varepsilon_1\eta_2^*-1)-2\eta_2^*}{\varepsilon_0(\varepsilon_1\eta_2-1)-2\eta_2}\, , \quad
    S^- = \xi_4^*\frac{\varepsilon_1\eta_2^*-1}{\varepsilon_1\eta_2-1}\, .
\end{equation}

\subsubsection{Unitary fidelity for distinguishable particles}\label{app:unitary_fidelity}
As mentioned in Sec. \ref{sec:ideal_fidelity}, the ideal fidelity of the CP gate for distinguishable plane waves is equal to the square modulus of the transmission coefficient $T$. Therefore, in order to achieve unitary fidelity with the distinguishable-particle architecture, one must tune the values of the interaction terms as to obtain $|T|^2=1$. This condition can readily be derived from the expressions in Appendix \ref{app:scattering}. {For example, the solutions for $C=1$, by setting $q_1=-q_2=q_r$ as in Fig. \ref{fig:fid_ideal_C01}, are obtained from the expression in Eq. \eqref{eq:t1}. Specifically, the square modulus of the transmission coefficient can be writte, after some manipulation, as:
\begin{align}
    &\left|T_0^{(C=1)}(q_r)\right|^2 =\frac{4\sin^2(q_r)}{\nu_1\zeta_1+4\cos^2(q_r)\varepsilon_0\varepsilon_1\mu_1+4\cos(q_r)\rho_1}\, ,
\end{align}
where we defined the following quantities:
\begin{align}
    \mu_1 =& \varepsilon_0\varepsilon_1-2\, , \\
    \nu_1 =& \varepsilon_1^2+1\, , \\
    \rho_1 =& (\mu_1+1)(\varepsilon_1\mu_1+\varepsilon_0)\, , \\
    \zeta_1 =& \mu_1^2+\varepsilon_0^2\, .
\end{align}
After setting $w_0=w_1$ (i.e. $\varepsilon_0 =\varepsilon_1$) as in Fig. \ref{fig:fid_ideal_C01}(b), one can find the solutions $\pm q_{r,1}$ and $\pm q_{r,2}$ for unitary fidelity mentioned in Sec. \ref{sec:ideal_fidelity} of the main text:
\begin{align}
    q_{r,1} =&\, -i\ln\left(w_0+2+\sqrt{w_0^2+4w_0-12}\right)\, , \\
    q_{r,2} =&\, -i\ln\left(w_0-2+\sqrt{w_0^2-4w_0-12}\right) \, .
\end{align}
Notice that $q_{r,1}$ is real only for $-6\leq w_0\leq 2$, while $q_{r,2}$ only for $-2\leq w_0\leq 6$. As for the case described in Fig. \ref{fig:fid_ideal_C2}}, we can fix the particles' momenta to $q_1=-q_2=-\pi/2$, and focus our analysis on the most general of the considered cases, i.e. $C=2$. Using Eq. \eqref{eq:t_c2} and after some manipulations of the formulas, one can write:
\begin{equation}\label{eq:t_squared}
    \left|T_0^{(C=2)}\left(\frac{\pi}{2}\right)\right|^2 = \frac{4}{[\zeta_2\varepsilon_0-2(\varepsilon_1\nu_2-\varepsilon_2)]^2+4}\, ,
\end{equation}
where we defined the following quantities:
\begin{align}
    \nu_2 =& \varepsilon_2^2+1\, , \\
    \zeta_2 =& (\varepsilon_1\varepsilon_2-1)^2+\varepsilon_1^2\, ,
\end{align}
and, coherently with the values of the momenta, we have that $\varepsilon_r = w_r/2$. By setting the quantity in Eq. \eqref{eq:t_squared} equal to 1, one obtains the following condition:
\begin{equation}\label{eq:unitary_fid_condition}
    w_0 = \frac{2(w_1\nu-w_2)}{(w_1w_2/4-1)^2+w_1^2/4}\, .
\end{equation}
The condition above provides, for each considered value of $w_2$, the equation of the dashed grey lines in Fig. \ref{fig:fid_ideal_C2}(a). Specifically, by setting $w_2=0$, one finds the locus of points with unitary fidelity for the nearest-neighbor case ($C=1$):
\begin{equation}
    w_0 = \frac{2w_1}{1+w_1^2/4}\, .
\end{equation}
Notice that the latter condition can be satisfied only for $|w_0|<2$, as is also evident from Fig. \ref{fig:fid_ideal_C2}(a). Additionally, this condition allows $|w_0|\geq |w_1|$ only for $|w_1|\leq 2$; otherwise, the potential is not monotonic with $r$. If we go a step further and also set $w_1=0$ (i.e. on-site interaction, $C=0$), we see that the only case for which we obtain unitary fidelity is $w_0=0$, that is, for no interaction at all. This is in complete agreement with the discussion of Fig. \ref{fig:fid_ideal_C01}(a) in the main text.\newline

By replacing the condition in Eq. \eqref{eq:unitary_fid_condition} into Eq. \eqref{eq:t_c2}, one can then find the optimal parameters to obtain an induced phase of $\theta = \pm\pi/2$. After some calculations, one obtains the following conditions:
\begin{align}
    w_0 =& \mp (2-w_2^2/2)\, , \\
    w_1 =& \frac{4}{w_2\mp2}\, ,
\end{align}
which correspond to the potential shapes reported in Fig. \ref{fig:fid_ideal_C2}(c). Notice that, under these conditions, the potential shape is monotonically decreasing with $r$ if and only if $0<w_2<\sqrt{5}-1$ for $\theta=-\pi/2$, and monotonically increasing if and only if $1-\sqrt{5}<w_2<0$ for $\theta=\pi/2$.

\subsection{Derivation of CP gate fidelities with a wavepacket}\label{app:finite_fidelity}
In order to evaluate the fidelity from Eq. \eqref{eq:fidelity_wp_dist}, it is useful to rewrite the relative-coordinate part of the two-particle wavefunction in momentum space:
\begin{equation}\label{eq:psi_rel_momentum}
    \langle k_r|\psi_{{\rm in},r}\rangle = \frac{1}{(2\pi\sigma_k^2)^{1/4}}e^{-\frac{(k_r+q_r)^2}{4\sigma_k^2}}e^{-ik_rr_0}
\end{equation}
In the case of distinguishable particles, the transmission coefficient $T_q(k_r) = |T_q(k_r)|e^{i\theta(k_r)}$ must be expanded by taking account of both its modulus and its phase. Specifically, we can Taylor expand the phase as:
\begin{equation}\label{eq:S_expansion}
    e^{i\theta(k_r)}\simeq \exp[i(\theta_0+\theta_1(k_r+q_r)+\theta_2(k_r+q_r)^2]\, ,
\end{equation}
where we defined $\theta_0=\theta(-q_r)$; $\theta_1=\partial_{k_r}\theta(k_r)|_{-q_r}$; and $\theta_2=1/2\,\partial^2_{k_r}\theta(k_r)|_{-q_r}$. Then, we expand the logarithmic magnitude as:
\begin{align}\label{eq:T_expansion}
    \ln|T_q(k_r)| \simeq l_0+l_1(k_r+q_r)+l_2(k_r+q_r)^2\, ,
\end{align}
with $l_0=\ln|T_q(-q_r)|$; $l_1=\partial_{k_r} \ln|T_q(k_r)|\big|_{-q_r}$; and $l_2=1/2\,\partial^2_{k_r} \ln|T_q(k_r)|\big|_{-q_r}$. Now, replacing Eqs. \eqref{eq:psi_rel_momentum}, \eqref{eq:S_expansion} and \eqref{eq:T_expansion} into Eq. \eqref{eq:fidelity_wp_dist}, we can write:
\begin{widetext}
\begin{align}\label{eq:fidelity_big_calculation}
    F \simeq& \left|\frac{1}{(2\pi\sigma_k^2)^{1/2}}\int_{-\infty}^{\infty}e^{-\frac{(k_r+q_r)^2}{2\sigma_k^2}}\,e^{(l_0+i\theta_0)}e^{l_1(k_r+q_r)}e^{-i\theta_1q_r}e^{(l_2+i\theta_2)(k_r+q_r)^2}\, dk\,\right|^2 \nonumber \\
    =& \frac{2\sigma^2}{\pi}\,e^{2l_0}\left|\int_{-\infty}^{\infty}e^{-(k_r+q_r)^2(2\sigma^2-l_2-i\theta_2)}e^{l_1(k_r+q_r)}\, dk\,\right|^2 \nonumber \\
    =& \frac{2\sigma^2}{\pi}\,e^{2l_0}\left|\sqrt{\frac{\pi}{(2\sigma^2-l_2-i\theta_2)}}\exp\left\{\frac{l_1^2}{4(2\sigma^2-l_2-i\theta_2)}\right\}\right|^2 \nonumber \\
    =& \sqrt{\frac{1}{(1-\xi)^2+\eta^2}}e^{2l_0}\left|\exp\left\{\frac{l_1^2}{8\sigma^2(1-\xi-i\eta)}\right\}\right|^2\, ,
\end{align}
\end{widetext}
where in the last step we defined $\xi=l_2/2\sigma^2$ and $\eta =\theta_2/2\sigma^2$. Notice that the Gaussian integral performed in the derivation is valid only for $1-\xi>0$, i.e. $l_2<2\sigma^2=(2\sigma_k^2)^{-1}$. We can then rewrite the complex denominator in the exponential as:
\begin{align}
    \frac{1}{1-\xi-i\eta}=\frac{1-\xi+i\eta}{(1-\xi)^2+\eta^2}\, ,
\end{align}
which we can then replace into the formula:
\begin{align}\label{eq:fidelity_dist}
    F\simeq&\frac{1}{\sqrt{(1-\xi)^2+\eta^2}}e^{2l_0}\left|\exp\left\{\frac{l_1^2(1-\xi+i\eta)}{8\sigma^2[(1-\xi)^2+\eta^2]}\right\}\right|^2 \nonumber \\
    =&\frac{|T_q(-q_r)|^2}{\sqrt{(1-\xi)^2+\eta^2}} \exp\left\{\frac{l_1^2(1-\xi)}{4\sigma^2[(1-\xi)^2+\eta^2]}\right\}\, .
\end{align}
We can now approximate by expanding the prefactor and the denominator in the exponential, keeping only the lowest-order terms in $\xi$ and $\eta$:
\begin{align}\label{eq:fid_appendix}
    \frac{1}{\sqrt{1+\xi^2-2\xi+\eta^2}}\simeq& 1-\frac{1}{2}(-2\xi+\eta^2)\simeq e^{\xi-\eta^2/2} \nonumber \\
    =&\, e^{l_2/2\sigma^2-\theta_2^2/8\sigma^4} = e^{2(l_2\sigma_k^2-\theta_2^2\sigma_k^4)}\,  \\
    \frac{1}{1+\xi^2-2\xi+\eta^2}\simeq& 1-(-2\xi+\eta^2)\, .
\end{align}
By replacing into Eq. \eqref{eq:fidelity_dist}, and again neglecting higher-order terms, we get:
\begin{align}\label{eq:fid_dist_appendix}
    F\simeq& |T_q(-q_r)|^2\exp[(l_1^2+2l_2)\sigma_k^2]\exp(-2\theta_2^2\sigma_k^4) \nonumber \\
    =& F_{\rm magn}F_{\rm phase}\, .
\end{align}
As we see, this defines two contributions in the fidelity variation due to finite-size effects, respectively because of the change in the magnitude or in the phase of $T(k_r)$. When $l_0,l_1,l_2=0$ (i.e. $|T_q(k)|\equiv1$) we find the result for a pure phase variation, which is:
\begin{equation}\label{eq:fidelity_dist_phase}
    F_{\rm phase}\simeq \exp(-2\theta_2^2\sigma_k^4)\, .
\end{equation}
At the same time, by fixing $\theta_2=0$ in Eq. \eqref{eq:fidelity_dist} one obtains the fidelity in case of pure magnitude variation:
\begin{equation}\label{eq:fidelity_dist_magn}
    F_{\rm magn}\simeq |T_q(-q_r)|^2\exp[(l_1^2+2l_2)\sigma_k^2]\, .
\end{equation}
Notice that, contrarily to the previous contribution, this one is possibly able to increase the fidelity, if either $l_1\neq 0$ or $l_2>0$.\newline

\begin{figure}[b!]
    \centering
    \includegraphics[width=\linewidth]{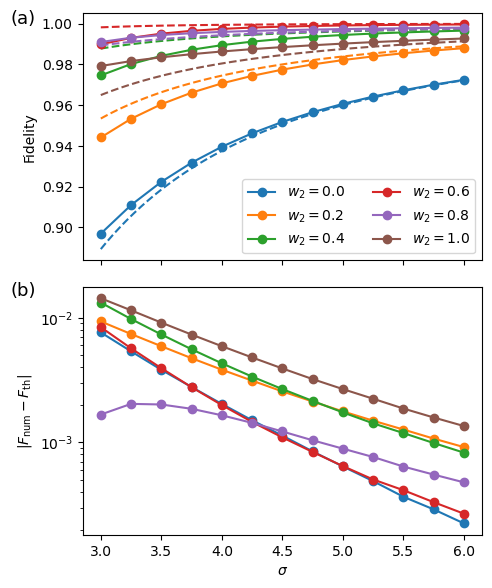}
    \caption{Fidelity of the distinguishable-particle CP gate as a function of the relative-coordinate wavepacket width $\sigma$, under the same gate conditions as in Fig. \ref{fig:fid_sigma6}. (a) Numerical values (continuous lines) and analytical approximation (dashed lines) of the gate fidelity. (b) Absolute difference between the numerical and analytical approaches.}
    \label{fig:app_fid}
\end{figure}

Fig. \ref{fig:app_fid} shows a comparison between the fidelity obtained from the numerical simulations described in the main text and the analytical approximation in Eq. \eqref{eq:fid_dist_appendix}. As expected, the discrepancy between the two quantities decreases as the width $\sigma$ increases and, at the same time, the width in momentum space $\sigma_k$ decreases. Specifically, depending on the shape of the potential and the choice of values for the interaction terms, the error of the formula can reach values as low as $2$e-4 for $\sigma=6$. Despite being this discrepancy relatively high with respect to the fidelity loss itself (see main text), the analytical formula from Eq. \eqref{eq:fid_dist_appendix} provides a good indication of the order of magnitude of the fidelity loss. As such, it is still useful for considerations about different gate implementations and their performance.\newline

\begin{figure}[t!]
    \centering
    \includegraphics[width=\linewidth]{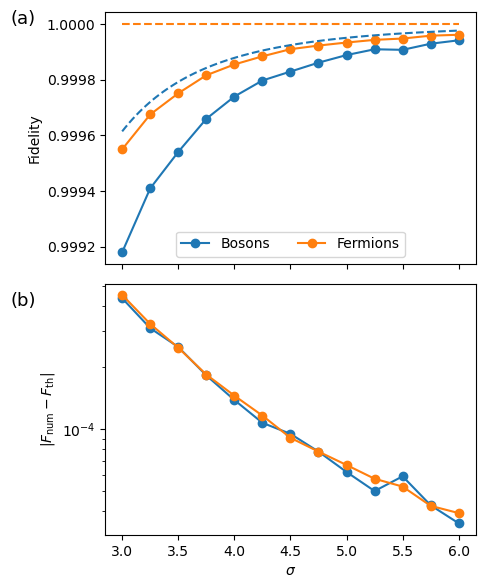}
    \caption{Fidelity of the bosonic and fermionic CP gate as a function of the relative-coordinate wavepacket width $\sigma$, respectively for on-site interaction ($w_0=4$) and first-neighbor interaction ($w_1=2$). (a) Numerical values (continuous lines) and analytical approximation (dashed lines) of the gate fidelity. (b) Absolute difference between the numerical and analytical approaches.}
    \label{fig:app_fid_indist}
\end{figure}

When considering indistinguishabale particles, the procedure to find the approximated fidelity is the same as the one illustrated above, only replacing $T\rightarrow S^\pm$. Since the magnitude of $S^\pm$ is identically 1, the formula for indistinguishable particle reduces to Eq. \eqref{eq:fidelity_dist_phase}, with $\theta_2$ now being the derivative of the phase of $S^\pm$. We show in Fig. \ref{fig:app_fid_indist} the comparison between the numerical fidelity from the simulations and the approximated formula. As above, the discrepancy decreases as $\sigma$ increases. Notice that, for the fermionic interaction potential chosen in the simulations ($w_1=2$), the phase increases linearly with $q_r$ [see Fig. \ref{fig:fid_ideal_C01}(d)], so that $\theta_2=0$ and the approximated fidelity from Eq. \eqref{eq:fidelity_dist_phase} identically zero. This implies that, only in this case, higher orders in $\sigma_k^6$ should be taken into account in the approximation. However, since the scope of this work does not require further analysis in these terms, we postpone it for future works.

\begin{figure}[t!]
    \centering
    \includegraphics[width=\linewidth]{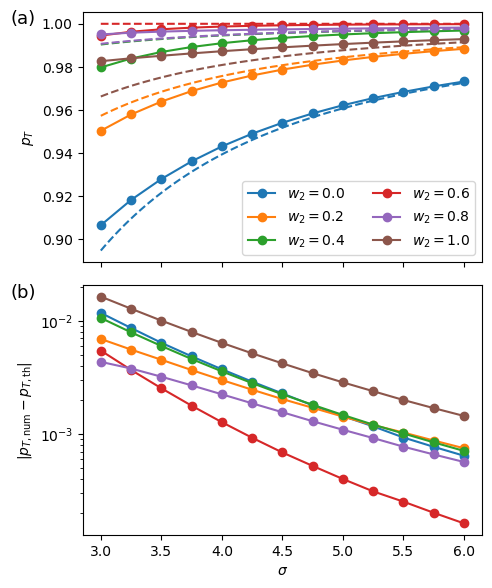}
    \caption{Transmission probability of the distinguishable-particle CP gate as a function of the relative-coordinate wavepacket width $\sigma$, under the same gate conditions as in Fig. \ref{fig:fid_sigma6}. (a) Numerical values (continuous lines) and analytical approximation (dashed lines) of the transmission probability. (b) Absolute difference between the numerical and analytical approaches.}
    \label{fig:app_pt}
\end{figure}

\subsubsection{Transmission probability for distinguishable particles}\label{app:finite_transmission}
While in the indistinguishable case total transmission always occurs after scattering, in the distinguishable case the presence of a finite $\sigma_k$ may modify the total transmission coefficient. Indeed, defining the transmission probability as:
\begin{align}
    p_T=&\langle\psi_{{\rm out},r}|\psi_{{\rm out},r}\rangle \nonumber \\
    =& \int_{-\infty}^\infty \langle\psi_{{\rm in},r}|k_r\rangle T_q^*(k_r)T_q(k_r)\langle k|\psi_{{\rm in},r}\rangle dk_r\nonumber \\
    =& \int_{-\infty}^\infty P(k_r)|T_q(k_r)|^2dk_r\, ,
\end{align}
once again it depends on the behaviour of $T_q(k_r)$ around the central momentum. Indeed, inserting Eqs. \eqref{eq:psi_rel_momentum} and \eqref{eq:T_expansion}, we get:
\begin{align}
    p_T=& \frac{1}{(2\pi\sigma_k^2)^{1/2}} \int_{-\infty}^\infty e^{-\frac{(k_r+q_r)^2}{2\sigma_k^2}}e^{2l_0+2l_1(k_r+q_r)+2l_2(k_r+q_r)^2}dk \nonumber \\
    =& \sqrt{\frac{2\sigma^2}{\pi}} e^{2l_0}\int_{-\infty}^\infty e^{-(k_r+q_r)^2(2\sigma^2-2l_2)}e^{2l_1(k_r+q_r)}dk \nonumber \\
    =& \sqrt{\frac{2\sigma^2}{\pi}} e^{2l_0} \sqrt{\frac{\pi}{(2\sigma^2-2l_2)}}e^{\frac{l_1^2}{2\sigma^2-2l_2}} \nonumber \\
    =& \frac{|T_q(-q_r)|^2}{\sqrt{1-2\xi}}e^{\frac{l_1^2}{2\sigma^2(1-2\xi)}}\, ,
\end{align}
where, as before, we defined $\xi=l_2/2\sigma^2$, and the Gaussian integral is valid only for $l_2<\sigma^2 = (2\sigma_k)^{-2}$. By expanding both the prefactor and the denominator at the exponential:
\begin{align}
    \frac{1}{\sqrt{1-2\xi}} \simeq& 1-\frac{1}{2}(-2\xi) \simeq e^{\xi} = e^{l_2/2\sigma^2} = e^{2l_2\sigma_k^2}\, , \\
    \frac{1}{1-2\xi} \simeq& 1-(-2\xi)\, , 
\end{align}
and neglecting higher-order terms, we get:
\begin{align}\label{eq:trans_dist}
    p_T \simeq |T_q(-q_r)|^2\exp[2(l_1^2+l_2)\sigma_k^2]\, ,
\end{align}
which, as before, is potentially higher than the transmission probability for delocalized walkers for $l_1\neq 0$ and $l_2>0$. Importantly, comparing Eqs. \eqref{eq:fidelity_dist_magn} and \eqref{eq:trans_dist}, we get:
\begin{align}
    p_T\simeq F_{\rm magn}\exp(l_1^2\sigma_k^2)\leq 1\, ,
\end{align}
so that, if the approximation stands, $F_{\rm magn}$ is always smaller or equal to $p_T$, which in turn is upper limited by $1$.

Fig. \ref{fig:app_pt} shows a comparison between the numerical value of $p_T$ obtained from the simulations and the approximated formula from Eq. \eqref{eq:trans_dist}. The behaviour is equivalent to that in Fig. \ref{fig:app_fid}.

\subsubsection{Fidelity of the roundabout gate}\label{app:roundabout_fidelity}
Despite single-qubit gates not being the focus of this work, we report the same results for fidelity loss for the roundabout gates shown in Fig. \ref{fig:fig1}(a), since they are a fundamental element of the architecture from Ref. \cite{asaka2023} for indistinguishable particles. The procedure is the same as that used in to find Eq. \eqref{eq:fid_appendix}. However, in this case we focus on the single-particle wavepacket, and the ideal and real outputs of the gate for particle $i$ entering from lead $u$ and exiting in lead $v$ are:
\begin{equation}
    |\psi_{{\rm ideal},i}\rangle = e^{i\phi_0}e^{i\phi_1\hat{k}_i}
|\psi_{{\rm in},i}\rangle\, ,
\end{equation}
\begin{equation}
    |\psi_{{\rm out},i}\rangle = S_{\rm R}(k_i)
|\psi_{{\rm in},i}\rangle\, ,
\end{equation}
where $S_{\rm R}(k_i)$ is relevant matrix element of the scattering matrix introduced in Ref. \cite{asaka2023} for the roundabout gate. In the formulas above we defined $\phi_0$ and $\phi_1$ as the argument of $S_{\rm R}(k)$ and its first derivative evaluated at the central wave vector $q_i$. As before, $\phi_1$ can be interpreted as an effective length for which the wavepacket is shifted when going through the gate. Using the same convention as in Ref. \cite{asaka2023}:
\begin{equation}
    S_{\rm R}(k_i) = S_{10}(k_i) = \pm\frac{2e^{7ik_i/2-i\pi/4}\cos\left(\frac{k_i}{2}+\frac{\pi}{4}\right)}{2-i\cot(k_i)}\, ,
\end{equation}
where $0$ and $1$ are respectively the input and output leads of the gate and the sign depends on the internal graph structure of the gate. By following the same steps used to find Eq. \eqref{eq:fidelity_wp_dist}, the roundabout-gate fidelity for a Gaussian wavepacket can therefore be written as
\begin{equation}\label{eq:fidelity_roundabout}
F_{\rm R} = \left|\int_{-\infty}^{+\infty}e^{-i\phi_1k_i}P(k_i)S_{\rm R}(k_i)\,dk_i\,\right|^2 \, ,
\end{equation}
where $P(k_i)=|\langle k_i|\psi_{{\rm in},i}\rangle|^2$ is the momentum probability distribution defined on the shape of the incoming wavepacket, as seen in Eq. \eqref{eq:psi_momentum}. Writing the scattering amplitude as:
\begin{equation}
S_{\rm R}(k_i)=|S_{\rm R}(k_i)|e^{i\phi(k_i)},
\end{equation}
allows us to expand its logarithmic magnitude and phase around the central momentum $q_i$, like we did before for $T_q(k_i)$ in Eqs. \eqref{eq:S_expansion} and \eqref{eq:T_expansion}:
\begin{align}
    \phi(k_i)&\simeq\phi_0+\phi_1(k_i-q_i)+\phi_2(k_i-q_i)^2\, , \\
    \ln |S_{ts}(k_i)| &\simeq m_0+m_1(k_i-q_i)+m_2(k_i-q_i)^2\, ,
\end{align}
with $\phi_2 = 1/2\ \partial_{k_i}^2\phi(k_i)|_{q_i}$; $m_0 = \ln|S_{\rm R}(q_i)|$; $m_1 = \partial_{k_i}\ln|S_{\rm R}(k_i)|\big|_{q_i}$; and $m_2 = 1/2\partial_{k_i}^2\ln|S_{\rm R}(k_i)|\big|_{q_i}$. By doing this and replacing in Eq. \eqref{eq:fidelity_roundabout} one obtains the same Gaussian integral as in Eq. \eqref{eq:fidelity_big_calculation}, only with $\sigma_k\rightarrow \tilde{\sigma}_k$ and $-q_r\rightarrow q_i$. This yields:
\begin{align}
    F_{\rm R}\simeq& |S_{\rm R}(q_i)|^2\exp[(m_1^2+2m_2)\tilde{\sigma}_k^2]\exp(-2\phi_2^2\tilde{\sigma}_k^4) \nonumber \\
    =& |S_{\rm R}(q_i)|^2\exp[2(m_1^2+2m_2)\sigma_k^2]\exp(-8\phi_2^2\sigma_k^4) \nonumber \\
    =& F_{\rm R,magn}F_{\rm R,phase}\, ,
\end{align}
where we used $\sigma_k = \tilde{\sigma}_k/\sqrt{2}$. Notice that, unlike the two-particle scattering phase considered in the main text for indistinguishable particles, the roundabout gate possesses a non-trivial momentum-dependent transmission amplitude. Consequently, both phase and magnitude variations contribute to the fidelity loss.

\subsubsection{Combined fidelity for the roundabout-assisted CP gate}\label{app:combined_fidelity}

In order to evaluate the complete fidelity of the setup illustrated in Fig. \ref{fig:fig1}(a), one should evaluate the fidelity of the entire process from the total scattering amplitude associated with the logical $|11\rangle$ branch of the circuit, as shown in Eq. \eqref{eq:tot_scattering_indist}. In order to do so, one must keep in mind that the single-particle roundabout gate and the two-particle interaction gate act on different momentum coordinates. Indeed, the roundabout scattering amplitudes are functions of the individual particle momenta, $S_{\rm R}(k_1)$ and $S_{\rm R}(k_2)$, whereas the two-particle scattering coefficient from Sec.~\ref{sec:indist_ptc} is naturally expressed in terms of the total and relative momenta, $k=k_1+k_2$ and $k_r=(k_1-k_2)/2$. The fidelity of the complete gate is therefore:
\begin{equation}\label{eq:fid_tot_int}
F_{\rm tot}
=
\left|
\iint 
P(k_1,k_2)\,
e^{-i\sum_i\Phi_i\delta k_i}
S_{\rm tot}(k_1,k_2)
dk_1dk_2\,\right|^2 ,
\end{equation}
where $P(k_1,k_2)$ is the joint momentum distribution of the incoming two-particle wavepacket, and $\Phi_i$ is the first derivative of the phase of the total scattering amplitude evaluated at the central momentum of particle $i$, with $\delta k_i = k_i-q_i$. Once again, the values of $\Phi_i$ can be interpreted as effective lengths. Expanding the logarithmic magnitude and phase of $S_{\rm tot}(k_1,k_2)$ around the central momenta $q_1$ and $q_2$, one gets:
\begin{align}
\label{eq:ln_Stot}
\ln |S_{\rm tot}|
&\simeq
L_0
+
\sum_i L_i \delta k_i
+
\sum_{ij}L_{ij}\delta k_i\delta k_j ,
\\
\label{eq:Phi_tot}
\Phi
&\simeq
\Phi_0
+
\sum_i \Phi_i \delta k_i
+
\sum_{ij}\Phi_{ij}\delta k_i\delta k_j ,
\end{align}
where we introduced the central magnitude and phase $L_0$ and $\Phi_0$, the first derivatives $L_i$ and $\Phi_i$, and the second derivatives $L_{ij}$ and $\Phi_{ij}$. At this point, one can repeat similar steps as before to obtain the fidelity of the complete gate. For convenience, let us introduce the matrices:
\begin{equation}
\mathbf{L} =
\begin{pmatrix}
L_{11} & L_{12} \\
L_{12} & L_{22}
\end{pmatrix},
\qquad
\mathbf{\Phi} =
\begin{pmatrix}
\Phi_{11} & \Phi_{12} \\
\Phi_{12} & \Phi_{22}
\end{pmatrix},
\end{equation}
together with the vectors $\mathbf{k}=(\delta k_1,\delta k_2)^T$ and $\mathbf{L}^{(1)}=(L_1,L_2)^T$. The symmetry relations $L_{ij}=L_{ji}$ and $\Phi_{ij}=\Phi_{ji}$ follow directly from the equality of mixed second derivatives. By replacing Eqs.~\eqref{eq:ln_Stot} and \eqref{eq:Phi_tot} into Eq.~\eqref{eq:fid_tot_int}, one obtains:
\begin{align}\label{eq:ftot_app}
&F_{\rm tot}
\simeq e^{2L_0} \times \nonumber \\
&\;
\left|
\iint
P(k_1,k_2)
\exp\!\left[
\mathbf{k}^{T}
(\mathbf{L}+i\mathbf{\Phi})
\mathbf{k}
+
\mathbf{L}^{(1),T}\!\cdot\!\mathbf{k}
\right]
dk_1dk_2
\right|^2 ,
\end{align}
The resulting integral is a two-dimensional Gaussian integral and can be evaluated analytically through the determinant and inverse of the corresponding quadratic form.\newline

An important difference with respect to the single-gate calculations is the presence of the mixed coefficients \(L_{12}\) and \(\Phi_{12}\), which generally do not allow for the factorization of the integral in Eq. \eqref{eq:ftot_app}. These quantities quantify correlations between the momentum components of the two walkers. Such terms are zero when considering only single-particle gates, as the amplitude and phase of each gate only acts on one momentum coordinate. 
%
Therefore, the mixed derivatives originate solely from the interaction region, and encode genuine two-particle correlations generated during the scattering process. At this point, since the magnitude $|S^{\pm}|$ of the interaction scattering is identically one, $L_{12}=L_{21}=0$. However, the same cannot be said for the scattering phase $\theta(q_r)$, so that:
\begin{equation}
    \Phi_{12}=\frac{1}{2}\frac{\partial^2 \theta(k_r)}{\partial k_1\partial k_2}\bigg|_{q_1,q_2} = \frac{1}{2} \frac{\partial k_r}{\partial k_1}\frac{\partial k_r}{\partial k_2}\frac{\partial^2\theta(k_r)}{\partial k_r^2}\bigg|_{-q_r} = -\frac{1}{4}\theta_2\, .
\end{equation}
Because this quantity is generally non-zero, a rigorous evaluation of the fidelity of the architecture shown in Fig.~\ref{fig:fig1}(a) requires the full two-dimensional Gaussian integration of Eq.~\eqref{eq:ftot_app}, taking into account the complete quadratic form of the total scattering phase. That said, when $|\theta_2|\sigma_k^2\ll1$, the correlated phase contribution remains perturbatively small over the momentum width of the incoming wavepacket with respect to the other contributions. Only in this regime, the total fidelity can be approximated by the product seen in Eq. \eqref{eq:ftot_prod}.


\bibliography{bib.bib}

\end{document}